\documentclass[graybox]{svmult}

\usepackage{mathptmx}       
\usepackage{helvet}         
\usepackage{courier}        
\usepackage{type1cm}        
\usepackage{makeidx}         
\usepackage{graphicx}        
\usepackage{multicol}        
\usepackage[bottom]{footmisc}
\newcommand{\BLGX}{BLG184.7\thinspace 133264}

\makeindex             

\begin{document}

\title*{First detection of period doubling in a BL~Herculis type star. Observations and theoretical models}
\titlerunning{Period doubling in BL~Herculis type star}
\author{R.~Smolec, I.~Soszy\'nski, P.~Moskalik, A.~Udalski, M.K.~Szyma\'nski, M.~Kubiak, G.~Pietrzy\'nski, \L{}.~Wyrzykowski, K.~Ulaczyk, R.~Poleski, S.~Koz\l{}owski and P.~Pietrukowicz}
\authorrunning{R.~Smolec, I.~Soszy\'nski, P.~Moskalik et al.}
\institute{R. Smolec \at Institute of Astronomy (IfA), University of Vienna, T\"urkenschanzstrasse 17, A-1180 Wien, Austria, \email{radek.smolec@univie.ac.at}
\and I.~Soszy\'nski, A.~Udalski, M.K.~Szyma\'nski, M.~Kubiak, G.~Pietrzy\'nski, \L{}.~Wyrzykowski, K.~Ulaczyk, R.~Poleski, S.~Koz\l{}owski and P.~Pietrukowicz \at Warsaw University Observatory, Al. Ujazdowskie 4, 00-478 Warszawa, Poland
\and P. Moskalik \at Copernicus Astronomical Centre, Bartycka 18, 00-716 Warszawa, Poland}
%
%
\maketitle

\abstract*{We report on the discovery of the first BL~Herculis star displaying 
period-doubling behaviour as predicted by the hydrodynamic models of Buchler
\& Moskalik \cite{BM92}. The star, with $P_0=2.4$\thinspace d, is located in 
the Galactic bulge and was discovered with OGLE-III photometry. We present 
new nonlinear convective models, which, together with recent evolutionary 
tracks, put constraints on the stellar parameters. In particular, we estimate 
the mass and metallicity of the object.}

\abstract{We report on the discovery of the first BL~Herculis star displaying 
period-doubling behaviour as predicted by the hydrodynamic models of Buchler
\& Moskalik \cite{BM92}. The star, with $P_0=2.4$\thinspace d, is located in 
the Galactic bulge and was discovered with OGLE-III photometry. We present 
new nonlinear convective models, which, together with recent evolutionary 
tracks, put constraints on the stellar parameters. In particular, we estimate 
the mass and metallicity of the object.}

\section{Introduction}

BL~Herculis stars are a subgroup of type II (or Population II) Cepheids, 
pulsating with periods between $1$ and $4$ days (see \cite{Wal} for a recent 
review). They show a singly-periodic large-amplitude light variation. Such 
behaviour was qualitatively reproduced with early nonlinear radiative models, 
e.g. Buchler \& Moskalik \cite{BM92, MB93}. In several of their models with 
periods between $2$ and $2.6$ days, Buchler \& Moskalik \cite{BM92} found 
period-doubling behaviour -- oscillations with periodic alternations of deep 
and shallow minima -- a phenomenon not observed in any BL~Her star at that time. 
Here we report on the discovery of the first BL~Her star clearly showing 
period-doubling behaviour.  We summarize the observations and new 
pulsation models for this star. A more detailed analysis, as well as discussion of another BL~Her star in which period-doubling behaviour is strongly suspected can be found in Smolec et al. \cite{SmolecEtal11}.

\section{Observations}

\begin{figure}[t]
\sidecaption[t]
\includegraphics[scale=.55]{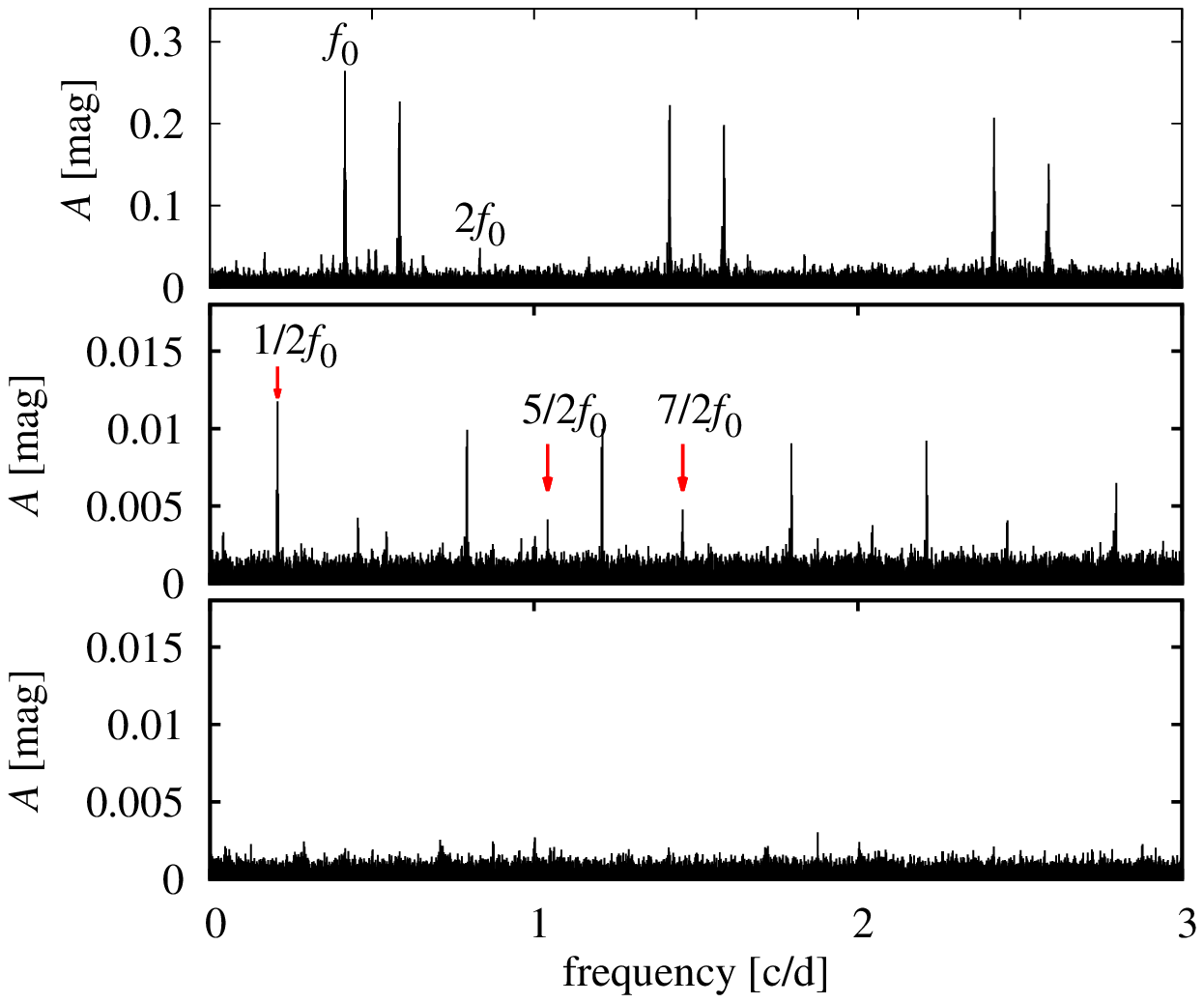}
\caption{Pre-whitening sequence for \BLGX. Upper panel: power spectrum of 
the original data. The spectrum is dominated by a pulsation frequency $f_0$ 
and its daily aliases. Middle panel: power spectrum after removing $f_0$ 
and its harmonics. The highest peak corresponds to a subharmonic frequency 
$\frac{1}{2}f_0$.  The daily aliases are also prominent.  Lower panel: 
power spectrum after removing $f_0$, its harmonics and its subharmonics.}
\label{fig.prewhitening}
\end{figure}

\begin{figure}[t]
\sidecaption
\includegraphics[scale=.55]{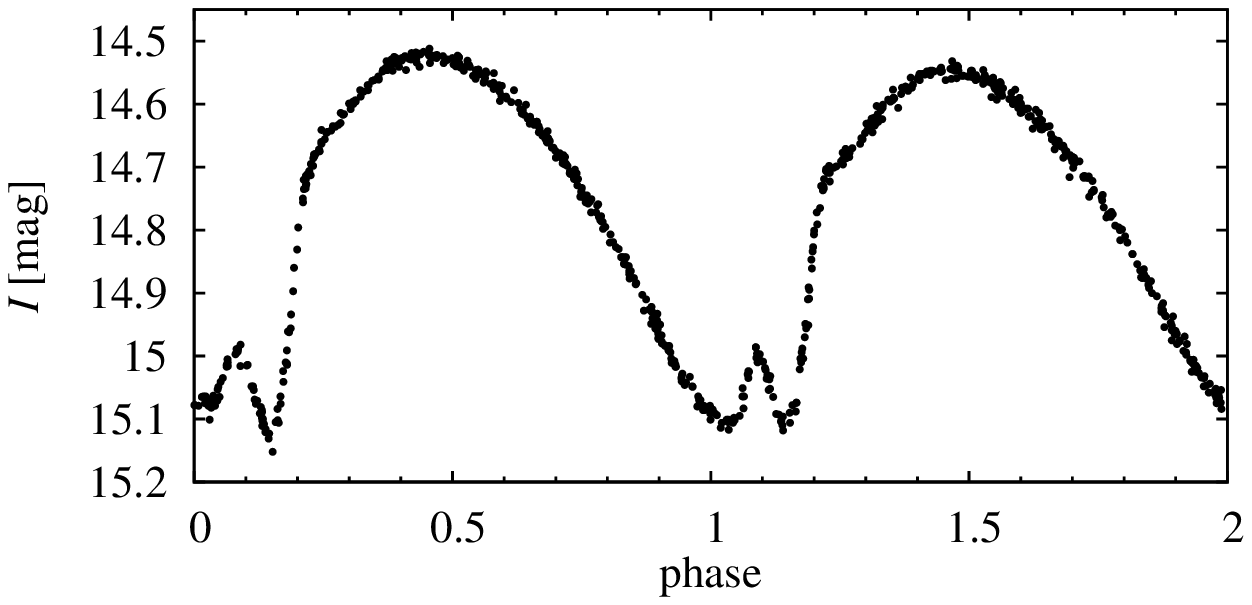}
\caption{Light curve of \BLGX\ phased with twice the pulsation period, $2P_0$.}
\label{fig.lc}
\end{figure}

The star of interest, \BLGX\ is in the Galactic bulge and was discovered
in data collected during the third phase of the Optical Gravitational Lensing 
Experiment \cite{OGLE}. It pulsates in the fundamental mode with a period of 
$P_0=2.4$\thinspace d. The {\it I}-band data for the star were analysed using 
standard consecutive pre-whitening technique.  Results are presented in 
Fig.~\ref{fig.prewhitening}. After removing the fundamental mode frequency, 
$f_0$, and its harmonics, additional signals are visible (middle panel of 
Fig.~\ref{fig.prewhitening}). The dominant peak is located at $\frac{1}{2}f_0$, 
which is a subharmonic of the primary frequency. Other subharmonics
($\frac{5}{2}f_0$, $\frac{7}{2}f_0$ and $\frac{9}{2}f_0$) are also present.
The presence of these frequencies in the power spectrum is a characteristic 
signature of period doubling.  In the time domain it means that the light 
curve repeats itself after two pulsation periods, instead of one.  This
gives rise to strictly periodic alternations (Fig.~\ref{fig.lc}).

\section{Theoretical models}

Using our nonlinear convective pulsation codes \cite{SM08}, we have computed 
several model sequences in order to model the period doubling in \BLGX\ and 
to constrain its parameters. Computational details are presented in 
\cite{SmolecEtal11}. Here we focus on presenting our best-fitting models 
and the resulting constraints on the parameters of \BLGX.

\begin{figure}[t]
\includegraphics[scale=.50]{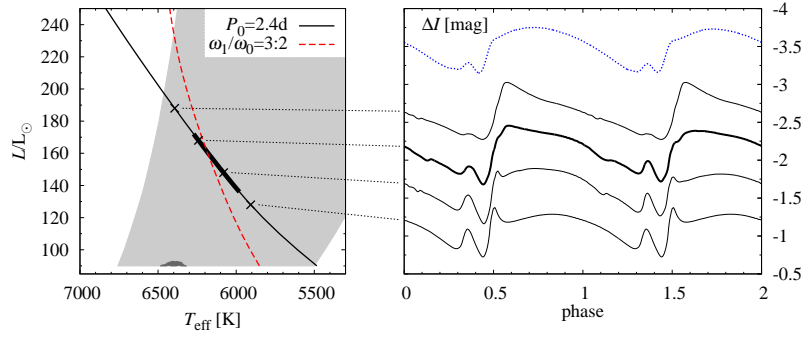}
\caption{The theoretical BL~Her instability strip in the HR diagram for 
$M=0.50{\rm M_\odot}$ and $Z=0.01$ (left panel). Nonlinear models are 
computed along a line of constant period, $P_0=2.4$\thinspace d.  The thick 
segment along this line indicates the period-doubling domain. Light 
curves for four selected models (marked with crosses) are displayed in the right panel (arbitrarily shifted in the vertical direction).  These are to be compared with the observed light curve (represented by a Fourier fit 
to the data) plotted at the top.}
\label{fig.hr}
\end{figure}

We have considered a grid of model masses ($M=0.50{\rm M_\odot}$, $0.55{\rm M_\odot}$, 
$0.60{\rm M_\odot}$ and $0.65{\rm M_\odot}$) and model metallicities 
($Z=0.01$, $0.001$ and $0.0001$). For each ($M$, $Z$) combination, an extensive 
grid of linear models was computed which covers the full BL~Her instability 
strip in the HR diagram (Fig.~\ref{fig.hr}). Next, a sequence of nonlinear 
models was computed along a line of constant period, $P_0=2.4$\thinspace d 
(right panel of Fig.~\ref{fig.hr}). Period-doubling behaviour was found 
over a range of luminosities.  In Fig.~\ref{fig.hr} these are shown by the
thick line for a particular ($M$, $Z$) combination. This domain correlates 
with the loci of the 3:2 half-integer resonance between the fundamental and 
first overtone modes, shown by the dashed line in Fig.~\ref{fig.hr}. As 
shown by  Moskalik \& Buchler \cite{MB90}, half-integer resonances are 
responsible for the period-doubling behaviour. Our detailed analysis \cite{SmolecEtal11} confirms that the 3:2 resonance indeed causes the period doubling behaviour in the BL~Her models, as Buchler \& Moskalik have already shown \cite{BM92}.

\begin{figure}[t]
\sidecaption[t]
\includegraphics[scale=.60]{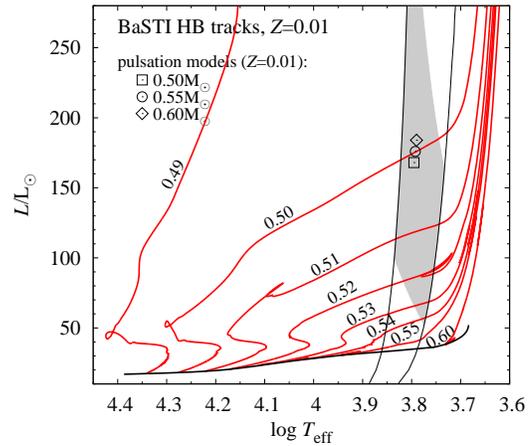}
\caption{The BaSTI \cite{BaSTI} horizontal branch evolutionary tracks for 
$Z=0.01$. Evolutionary tracks start at the Zero-Age Horizontal Branch (solid, 
horizontally running line). Each track is labeled with the corresponding 
model mass. Over-plotted are the edges of the instability strip, with the 
shaded area indicating the BL~Her domain with fundamental mode periods 
between 1\thinspace d  and 4\thinspace d. Pulsation models are plotted with 
different symbols.}
\label{fig.evol}
\end{figure}

The model light curves were compared with the observations. Our best-fitting
models closely match the pulsation amplitude and also match the amplitude of
the alternations (i.e., the amplitude of the subharmonic peak in the frequency 
spectrum) of \BLGX. A good match is possible only for models with the
highest metallicity, $Z=0.01$.  For lower metallicities, the pulsation 
amplitudes are lower than observed and, in addition, the amplitudes of the
alternations are much larger than observed.  This result is independent of the 
adopted values of the convective parameters, as analysed in detail in 
\cite{SmolecEtal11}. The three best-fitting models, all with $Z=0.01$, have 
different masses: $M=0.50{\rm M_\odot}$, $0.55{\rm M_\odot}$ and $0.60{\rm M_\odot}$. 
Based on pulsation computations alone, we cannot decide on the best model. 
However, since all these models fall roughly in the same place in the HR 
diagram (see Fig.~\ref{fig.evol}), evolutionary tracks may provide further 
constraints.

In Fig.~\ref{fig.evol} we plot horizontal branch evolutionary tracks 
from the BaSTI database corresponding to $Z=0.01$ \cite{BaSTI}. It is clear 
that only our least massive model ($M=0.50{\rm M_\odot}$) fits the evolutionary 
scenario. The tracks for larger masses run well beyond the instability strip. 
The light curve of the best model is shown in Fig.~\ref{fig.hr} (second model 
light curve from top, plotted with a thick line).

\begin{acknowledgement}
RS is supported by the Austrian FWF grant No. AP 21205-N16. Support from 
\"Osterreichische Forschungsgemeinschaft (Projekt 06/12308) is greatly 
acknowledged. The research leading to these results has also received 
funding from the European Research Council under the European Community's 
Seventh Framework Programme (FP7/2007-2013)/ERC grant agreement no. 246678.
\end{acknowledgement}

\end{document}